\documentclass[twocolumn,showpacs,aps,prl,superscriptaddress,xspace]{revtex4}

\usepackage{graphicx}
\usepackage{dcolumn}
\usepackage{amsmath}
\usepackage{amssymb}
\usepackage{epsfig}
\usepackage{psfrag}

\input pubboard/babarsym

\newcommand{\BABARPubYear}    {07}
\newcommand{\BABARPubNumber}  {004}
\newcommand{\SLACPubNumber} {12675}

\newcommand{\mmiss}{\ensuremath{m_\mathrm{miss}^2}\xspace}
\newcommand{\ds}{\ensuremath{D^{(*)}}\xspace}
\newcommand{\eextra}{\ensuremath{E_\mathrm{extra}}\xspace}
\newcommand{\btag}{\ensuremath{B_\mathrm{tag}}\xspace}
\newcommand{\pstarl}{\ensuremath{|{\bf p}^*_\ell|}\xspace}
\newcommand{\gevccnosp}{\ensuremath{{\mathrm{Ge\kern -0.1em V\!/}c^2}}\xspace}

\begin{document}

\preprint{\babar-PUB-\BABARPubYear/\BABARPubNumber} 
\preprint{SLAC-PUB-\SLACPubNumber} 

\begin{flushleft}
\babar-CONF-\BABARPubYear/\BABARPubNumber\\
SLAC-PUB-\SLACPubNumber\\
%hep-ex/\LANLNumber\\[10mm]
\end{flushleft}

\title{
{\large \bf
Measurement of the Semileptonic Decays {\boldmath $B\to D\taum\nutb$} and 
{\boldmath $B\to\Dstar\taum\nutb$}} 
}

%% author list as of 06-Jun-2007 (572 authors)
%
\author{B.~Aubert}
\author{M.~Bona}
\author{D.~Boutigny}
\author{Y.~Karyotakis}
\author{J.~P.~Lees}
\author{V.~Poireau}
\author{X.~Prudent}
\author{V.~Tisserand}
\author{A.~Zghiche}
\affiliation{Laboratoire de Physique des Particules, IN2P3/CNRS et Universit\'e de Savoie, F-74941 Annecy-Le-Vieux, France }
\author{J.~Garra~Tico}
\author{E.~Grauges}
\affiliation{Universitat de Barcelona, Facultat de Fisica, Departament ECM, E-08028 Barcelona, Spain }
% ...
\collaboration{The \babar\ Collaboration}
\noaffiliation

\date{\today}

\begin{abstract}
We present preliminary measurements of branching fractions for the semileptonic decays 
$B\to D\taum\nutb$ and $B\to\Dstar\taum\nutb$, which
are potentially sensitive to non--Standard Model amplitudes.
The data sample comprises $232\times 10^6$ 
$\Upsilon(4S)\to\BB$ decays collected with the \babar\ detector at the \pep2 \epem
storage ring. We obtain
${\cal B}(\Bm\to\Dz\taum\nutb)=(0.63\pm 0.38\pm 0.10\pm 0.06)\%$, 
${\cal B}(\Bm\to\Dstarz\taum\nutb)=(2.35\pm 0.49\pm 0.22\pm 0.18)\%$,
${\cal B}(\Bzb\to\Dp\taum\nutb)=(1.03\pm 0.35\pm 0.14\pm 0.10)\%$, 
and ${\cal B}(\Bzb\to\Dstarp\taum\nutb)=(1.15\pm 0.33\pm 0.04\pm 0.04)\%$, where the uncertainties
are statistical, systematic, and normalization, respectively.
By combining \Bm and \Bzb results, we also obtain the branching fractions
${\cal B}(B\to D\taum\nutb)=(0.90\pm 0.26\pm 0.11\pm 0.06)\%$ and
${\cal B}(B\to D^*\taum\nutb)=(1.81\pm 0.33\pm 0.11\pm 0.06)\%$ (quoted for
the \Bm lifetime), with significances of $3.5\sigma$ and $6.2\sigma$.

\vspace{3mm}

{\it Submitted to the 2007 Europhysics Conference on High Energy Physics,
Manchester, England.}
\end{abstract}


\maketitle

%%% Introduction
Semileptonic decays of $B$ mesons to the $\tau$ 
lepton---the heaviest of the three charged leptons---provide 
a new source of 
information on Standard Model (SM) 
processes~\cite{KornerSchuler,Falk_etal,HuangAndKim}, 
as well as a new window on physics beyond 
the SM~\cite{GrzadkowskiAndHou,Tanaka,KiersAndSoni,Itoh,ChenAndGeng}.
In the SM, semileptonic decays occur at tree level and are mediated by the $W^-$ boson, but
the large mass of the $\tau$ lepton provides sensitivity to 
additional amplitudes, such as those mediated by a charged Higgs 
boson, $H^-$. Experimentally, $b\to c\tau^-\nutb$ decays are 
challenging to study because the 
final state contains not just one, but two or three neutrinos.

%% Paragraph on theory predictions
Branching fractions for semileptonic $B$ decays to $\tau$ leptons are predicted to be
smaller than those for $\ell=e,\ \mu$~\cite{Ell}, but are still substantial compared to most hadronic
$B$ decays. A recent SM-based calculation~\cite{ChenAndGeng} predicts 
${\cal B}(\Bzb\to\Dp\taum\nutb)=(0.69\pm0.04)\%$ and
${\cal B}(\Bzb\to\Dstarp\taum\nutb)=(1.41\pm0.07)\%$; 
an inclusive calculation~\cite{Falk_etal} gives 
${\cal B}(B\to X_c\taum\nutb)=(2.3\pm0.25)\%$, where $X_c$ represents all
final states resulting from the $b\to c$ transition. 
Calculations~\cite{GrzadkowskiAndHou,Tanaka,KiersAndSoni,Itoh,ChenAndGeng} 
in supersymmetric models show that substantial
departures from the SM decay rate could occur for
${\cal B}(B\to D\taum\nutb)$, but that those for  
${\cal B}(B\to\Dstar\taum\nutb)$ are expected to be smaller. 
The interference with the SM amplitude can be constructive or destructive, depending on
the value of $(\tan\beta)/m_H$, where $\tan\beta$ is the ratio of the vacuum
expectation values for the two Higgs doublets and $m_H$ is the $H^-$ mass.

Theoretical predictions for semileptonic decays to exclusive final states 
require knowledge of the form factors, which parametrize the hadronic current
as a function of $q^2=(p_B-p_{\ds})^2.$ For light leptons ($e$, $\mu$), there is effectively
one form factor for $B\to D\ellm\nulb$, while there are 
three for $B\to\Dstar\ellm\nulb$. If a $\tau$ lepton is produced instead,
one additional form factor enters in each mode. The form factors for $B\to\ds\ellm\nulb$ 
decays involving the light leptons have been measured~\cite{FF} and have been
discussed extensively in the theoretical literature.
Heavy-quark-symmetry (HQS) relations~\cite{IsgurWise} allow one to express
the two additional form factors for $B\to\ds\taum\nutb$ in terms of the
form factors measurable from decays with the light leptons. With sufficient data, one could
probe the additional form factors and test the HQS relations.

%% Paragraph on previous meaurements
The first measurements of semileptonic $b$-hadron decays to $\tau$ leptons were performed by 
the LEP experiments~\cite{LEP} operating at the \Z resonance,
yielding an average~\cite{PDG} branching fraction
${\cal B}(b_{\rm had}\to X\taum\nutb)=(2.48\pm 0.26)\%$, where $b_{\rm had}$ 
represents the mixture of
$b$-hadrons produced in $\Z\to\bbbar$ decay.

We determine branching fractions of
four exclusive decay modes~\cite{ChargeConjugate}: $\Bm\to\Dz\taum\nutb$,
$\Bm\to\Dstarz\taum\nutb$, $\Bzb\to\Dp\taum\nutb$, and $\Bzb\to\Dstarp\taum\nutb$,
each of which is measured relative to the corresponding $e$ and $\mu$
modes. To reconstruct the $\tau$, we use the decays
$\taum\to\en\nueb\nut$ and $\taum\to\mun\numb\nut$, which are experimentally most
accessible. The main challenge of the
measurement is to separate $B\to\ds\taum\nutb$ decays, which have
three neutrinos, from $B\to\ds\ellm\nulb$ decays, which have
the same observable final-state particles but only one neutrino.

%% Paragraph on data sample and BaBar detector
We analyze data collected with the \babar\ detector~\cite{BABARNIM},
at the \pep2 \epem storage ring at 
the Stanford Linear Accelerator Center. The data sample used in the analysis 
comprises $208.9\ \invfb$ recorded on the $\FourS$
resonance, yielding $232\times 10^6\ \BB$ decays.
The measurement uses all of the major detector subsystems: 
a charged-particle tracking system consisting of a
5-layer silicon vertex tracker and a 40-layer He-gas drift chamber (DCH); a quartz-bar Cherenkov 
particle-identification system; a CsI(Tl) crystal electromagnetic calorimeter for
electron identification and photon energy measurement; a 1.5 T
superconducting magnet; and a muon identification system in the magnet flux return.

%%% Analysis overview and strategy

The analysis strategy is to reconstruct the decays of both $B$ mesons in the $\FourS\to\BB$ 
event, providing powerful constraints on unobserved particles. One $B$ 
meson, denoted \btag, is fully reconstructed in a purely hadronic decay chain. The remaining
charged tracks and photons are required to be consistent with the products of a $b\to c$
semileptonic $B$ decay: a hadronic system, either a $D$ or \Dstar meson,
and a lepton ($e$ or $\mu$), either primary or from
$\taum\to\ellm\nulb\nut$.
Using the known total four-momentum 
of the $\epem$ collision, we calculate $p_\mathrm{miss}=[p(\epem)-p_{\rm tag}-p_{\ds}-p_{\ell}]$ 
recoiling against the 
observed $\btag+\ds\ell$ system. A large peak at zero in $\mmiss= p_\mathrm{miss}^2$ corresponds 
to semileptonic decays with one neutrino, whereas signal events form a broad tail out to 
$\mmiss\sim 8\ (\gevccnosp)^2$. To separate signal and background events, we 
perform a fit to the joint distribution of \mmiss and the lepton 
momentum (\pstarl), in the rest frame of the $B$ meson.
In a signal event, the observed lepton is the daughter of the $\tau$ 
and typically has a soft spectrum; for most background events, this lepton 
typically has higher momentum.

%%% Event selection
We reconstruct \btag candidates~\cite{BtagAlgorithm} in 1114 final states
$\btag\to\ds Y^\pm$.
Tag-side \ds candidates are reconstructed in 21 decay chains, and the
$Y^\pm$ system can consist
of up to six light hadrons (\pipm, \piz, \Kpm, or \KS). \btag candidates
are identified using two kinematic variables, $\mes=\sqrt{s/4-|{\bf p}_\mathrm{tag}|^2}$
and $\DeltaE=E_\mathrm{tag}-\sqrt{s}/2$, where $\sqrt{s}$ is the total \epem
energy; $|{\bf p}_\mathrm{tag}|$ is the magnitude of the \btag momentum; and $E_\mathrm{tag}$ is
the \btag energy, all defined in the \epem center-of-mass frame. We require
$\mes>5.27\ \gevcc$ and $|\DeltaE|<72\ \mev$, corresponding
to $4\sigma$ (standard deviations). We reconstruct \btag candidates with an
efficiency of approximately $0.3\%$ to $0.5\%$.

For the $B$ meson decaying semileptonically, we reconstruct \ds candidates in the modes $\Dz\to\Km\pip$,
$\Km\pip\piz$, $\Km\pip\pipi$, $\KS\pipi$; $\Dp\to\Km\pip\pip$,
$\Km\pip\pip\piz$, $\KS\pip$, $\Km\Kp\pip$; $\Dstarz\to\Dz\piz$, $\Dz\gamma$;
and $\Dstarp\to\Dz\pip$, $\Dp\piz$. $D$ (\Dstar) candidates are selected within
$4\sigma$ of the $D$ mass ($\Dstar-D$ mass difference), with $\sigma$
typically $5$--$10\ \mevcc$ ($1$--$2\ \mevcc$).
To ensure well-measured momenta, identified electron and muon tracks are
required to have at least 12 hits in the drift chamber 
and not to be near the acceptance edges.
Electron candidates must have lab-frame momentum $|{\bf p}_e|>0.3 \gevc$;
muon candidates must have an appropriate 
signature in the muon detector system, effectively requiring  
$|{\bf p}_{\mu}|\gtrapprox 0.6 \gevc$. 
The energy of electron candidates is corrected for bremsstrahlung energy loss
if photons are found close to the electron direction. 

We require that there be no charged tracks not associated with
the \btag, \ds, or $\ell$ candidates. We compute \eextra, the sum of the
energies of all photon candidates not associated with the \btag, \ds, or
$\ell$ candidates, and we require $\eextra<150$--$300\ \mev$, depending
on the \ds channel. We suppress hadronic events and combinatoric
backgrounds by requiring $|{\bf p}_\mathrm{miss}|>200\ \mevc$ and $q^2>4\ (\gevccnosp)^2$.
If multiple candidates pass this selection, we select the
candidate with the lowest value of \eextra. To improve the \mmiss resolution,
we perform a kinematic fit to the event, constraining particle masses
to known values and requiring tracks from $B$, $D$, and \KS
mesons to originate from appropriate common vertices.

\begin{figure}[tb!]
\includegraphics[width=3.1in]{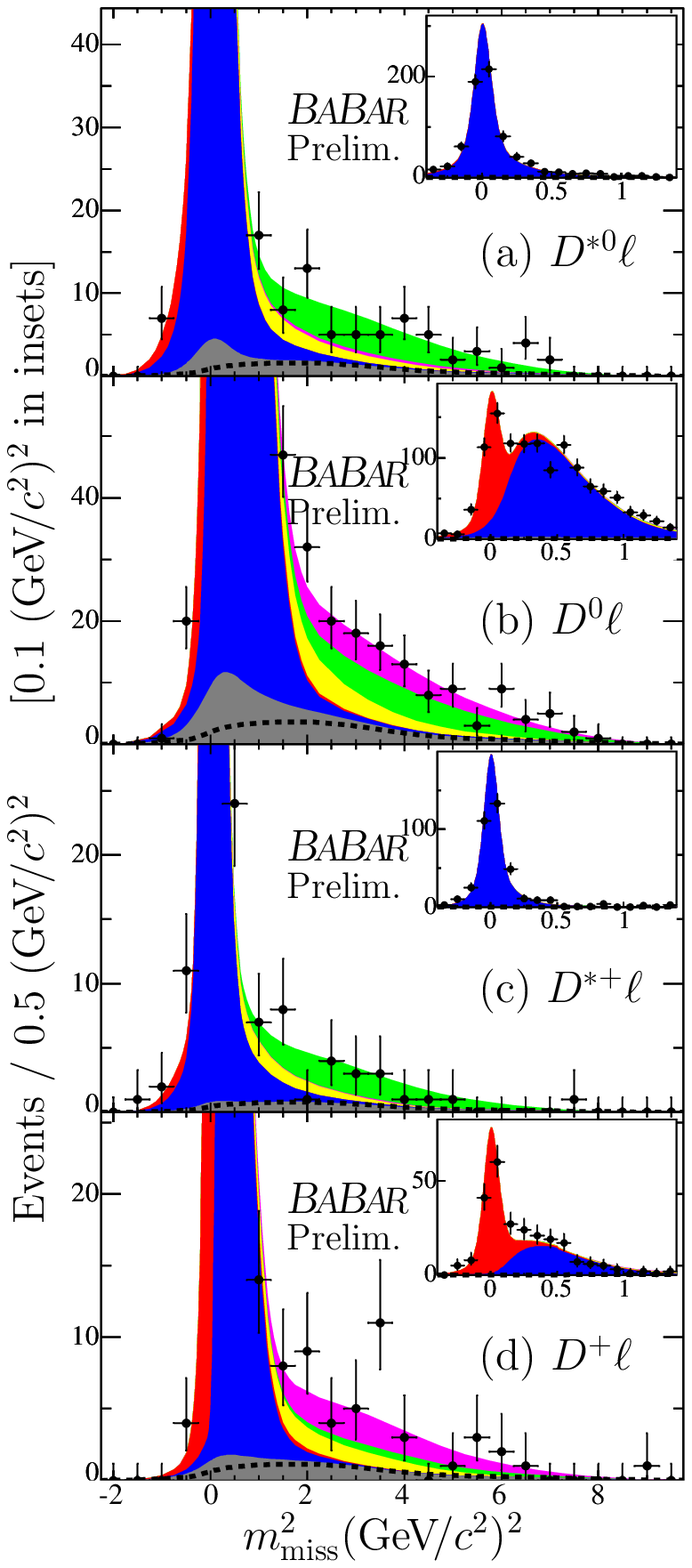}
\caption {Distributions of events and fit projections in \mmiss for the four signal channels:
$\Dstarz\ell$, $\Dz\ell$, $\Dstarp\ell$, and $\Dp\ell$. (The fit shown incorporates the
\Bm--\Bzb constraints.)
The normalization region $\mmiss\sim 0$ is shown as an inset in each figure. 
The fit components are: $\Dstar\taum\nutb$ (green), $D\taum\nutb$ (purple),
$\Dstar\ellm\nulb$ (blue), $D\ellm\nulb$ (red), $D^{**}(\ellm/\taum)\nub$ (yellow), combinatoric (grey,
below dashed line), charge crossfeed (grey, above dashed line).
}\label{fig:fit}
\end{figure}

Figure~\ref{fig:fit} shows the distributions of \mmiss for the
four $\ds\ell$ channels, along with the projections 
of the maximum likelihood fit to be discussed below. The large peaks at $\mmiss \approx 0$ 
are mainly due to $B\to\ds\ellm\nulb$, which serve as normalization modes.
The structure of this 
background is shown in the inset figures, which expand the region 
$-0.4<\mmiss< 1.4\ (\gevccnosp)^2$.
$B\to\Dstar\ellm\nulb$ background is the dominant feature in
the two $\Dstar\ell$ channels (Figs.~\ref{fig:fit}a, c);
the two $D\ell$ channels (Figs.~\ref{fig:fit}b, d) are dominated
by $B\to D\ellm\nulb$ decays but also include substantial feed-down
contributions from true \Dstar mesons where the low-momentum \piz
or photon from $\Dstar\to D\piz/\gamma$ is not reconstructed.
This feed-down is clearly visible for $B\to\Dstar\ellm\nulb$ background,
but affects $B\to\Dstar\taum\nutb$ signal similarly, and both
feed-down components (as well as smaller feed-up contributions from
$B\to D(\ellm/\taum)\nub$ into the $\Dstar\ell$ channels) are
included in the fit.
Other sources of background include
$B\to D^{**}(\ellm/\taum)\nub$ events (here $D^{**}$
represents charm resonances heavier than the $\Dstar(2010)$, as well as
non-resonant $\ds n\pi$ systems); charge-crossfeed (which occurs when
a $B\to\ds\ellm\nulb$ event is reconstructed with the wrong charge for the
\btag and \ds meson, typically because a low-momentum $\pipm$ is swapped between
the \btag and the \ds); and combinatoric background. This last background is dominated by
hadronic $B$ decays such as $B\to\ds D_s^{(*)}$, in which one 
of the charm mesons produces a secondary lepton, including $\tau$ leptons from 
$D_s$ decay. 

To constrain background from $B\to D^{**}(\ellm/\taum)\nub$ decays, we use
four control samples (one for each signal channel) 
in which an extra \piz meson is observed. Most of the $D^{**}$ background in the
signal channels occurs when the \piz from $D^{**}\to\ds\piz$ is not
reconstructed, so these control samples provide a good normalization of
the background source. $D^{**}$ decays in which a $\pipm$ is lost do not have the correct
charge correlation between the \btag and \ds, and decays with two missing charged
pions are rare. The feed-down probabilities for the $D^{**}(\ellm/\taum)\nub$
background are determined from simulation, with uncertainties in the
$D^{**}$ content treated as a systematic error. However, the control samples 
reduce our sensitivity to the details of this model.

%%% The fit
We perform a relative measurement, extracting both signal $B\to\ds\taum\nutb$
and normalization $B\to\ds\ellm\nulb$ yields from the fit to obtain the 
four branching ratios $R(\Dz)$, $R(\Dp)$, $R(\Dstarz)$, and $R(\Dstarp)$ where,
for example, $R(\Dstarz)\equiv{\cal B}(\Bm\to\Dstarz\taum\nutb)/{\cal B}(\Bm\to\Dstarz\ellm\nulb)$. 
Here, $\ell$ represents only one of $e$ or $\mu$; however, both light lepton species
contribute statistically to the denominator. Signal and background yields are extracted 
using an extended, unbinned maximum likelihood fit to the 
joint (\mmiss, \pstarl) distribution. The 18-parameter fit
is performed simultaneously in the four signal channels and the four $D^{**}$
control samples. In each of the four
signal channels, we describe the data as the sum of seven components (shown in Fig.~\ref{fig:fit}):
$D\taum\nutb$, $\Dstar\taum\nutb$, $D\ellm\nulb$, $\Dstar\ellm\nulb$, $D^{**}(\ellm/\taum)\nub$,
charge crossfeed, and combinatoric background. The four $D^{**}$ control samples
are described as the sum of five components: $D^{**}(\ellm/\taum)\nub$, $D\ellm\nulb$,
$\Dstar\ellm\nulb$, charge crossfeed, and combinatoric background. 
Probability distribution functions (PDFs) are primarily
determined from simulated event samples; however, the parameters describing
the dominant feed-down component---$\Dstar$ feed-down into 
the $D\ell$ channels---are determined directly by the fit.

%%% Results

We perform two fits, one in
which all four signal yields are allowed to float independently, and a second,
\Bm--\Bzb combined fit, in which we constrain~\cite{Isospin} 
$R(\Dp)=R(\Dz)$ and $R(\Dstarp)=R(\Dstarz)$.
The fit results are summarized in Table~\ref{tab:results}, 
and the \mmiss projections of the constrained fit are shown in Fig.~\ref{fig:fit}.

\begin{table*}
\caption{Preliminary results from fits to data and associated uncertainties: the columns are
the signal yield ($N_\mathrm{sig}$), the yield of normalization $B\to\ds\ellm\nulb$ events ($N_\mathrm{norm}$),
the ratio of signal and normalization mode efficiencies ($\varepsilon_\mathrm{sig}/\varepsilon_\mathrm{norm}$),
the relative systematic error due to the fit yields ($(\Delta R/R)_\mathrm{fit}$), the
relative systematic error due to the efficiency ratios ($(\Delta R/R)_\varepsilon$),
the branching fraction relative to the normalization mode ($R$), the absolute
branching fraction ($\cal B$), and the total and statistical signal significances
($\sigma_\mathrm{tot}$ and $\sigma_\mathrm{stat}$).
The first two errors on $R$
and $\cal B$ are statistical and systematic, respectively; the third error on $\cal B$
represents the uncertainty on the normalization mode~\cite{NormBF}.
The last two rows show the results of the fit with the \Bm--\Bzb constraint applied,
where $\cal B$ is expressed for the \Bm.
}
\label{tab:results}
\begin{tabular}{l l r@{$\pm$}l r@{$\pm$}l c c c r@{$\pm$}r@{$\pm$}r l@{$\pm$}l@{$\pm$}l@{$\pm$}l l}\\ \hline\hline
\multicolumn{2}{l}{Mode} & \multicolumn{2}{c}{$N_\mathrm{sig}$} &
  \multicolumn{2}{c}{$N_\mathrm{norm}$} &
  \multicolumn{1}{c}{$\varepsilon_\mathrm{sig}/\varepsilon_\mathrm{norm}$} &
  \multicolumn{1}{c}{$(\Delta R/R)_\mathrm{fit}$} &
  \multicolumn{1}{c}{$(\Delta R/R)_\varepsilon$} &
  \multicolumn{3}{c}{$R$} &
  \multicolumn{4}{c}{$\cal B$} & \multicolumn{1}{c}{$\sigma_\mathrm{tot}$} \\
\multicolumn{6}{c}{} &
  \multicolumn{1}{c}{} & \multicolumn{1}{c}{[\%]} &
  \multicolumn{1}{c}{[\%]} &
  \multicolumn{3}{c}{[\%]} & \multicolumn{4}{c}{[\%]} &
  \multicolumn{1}{c}{$(\sigma_\mathrm{stat})$} \\ \hline
\Bm & $\to\Dz\taum\nutb$ & $33.1$ & $ 19.6$ & $346.7$ & $23.0$ & $1.85$ & $15.2$ & $1.5$ &
  $29.5$ & $17.4$ & $4.5$ & $0.63$ & $0.38$ & $0.10$ & $0.06$ & $1.7\ (1.7)$ \\
\Bm & $\to\Dstarz\taum\nutb$ & $95.9$ & $19.8$ & $1628.6$ & $63.5$ & $0.98$ & $9.4$ & $1.4$ &
  $36.2$ & $7.5$ & $3.4$ & $2.35$ & $0.49$ & $0.22$ & $0.18$ & $5.3\ (5.8)$ \\
\Bzb & $\to\Dp\taum\nutb$ & $23.0$ & $7.9$ & $149.9$ & $13.3$ & $1.83$ & $13.4$ & $1.7$ &
  $48.6$ & $16.7$ & $6.6$ & $1.03$ & $0.35$ & $0.14$ & $0.10$ & $3.3\ (3.5)$ \\
\Bzb & $\to\Dstarp\taum\nutb$ & $16.2$ & $7.3$ & $481.8$ & $25.5$ & $0.93$ & $3.4$ & $1.5$ &
  $21.4$ & $9.7$ & $0.8$ & $1.15$ & $0.52$ & $0.04$ & $0.04$ & $2.7\ (2.7)$ \\ \hline
$B$ & $\to D\taum\nutb$ & $64.9$ & $19.1$ & $496.3$ & $26.4$ & $1.85$ & $12.0$ & $1.3$ &
  $40.7$ & $12.0$ & $4.9$ & $0.90$ & $0.26$ & $0.11$ & $0.06$ & $3.5\ (3.8)$ \\
$B$ & $\to\Dstar\taum\nutb$ & $105.3$ & $19.4$ & $2109.4$ & $ 68.0$ & $0.93$ & $5.7$ & $1.2$ &
  $31.0$ & $5.7$ & $1.8$ & $1.81$ & $0.33$ & $0.11$ & $0.06$ & $6.2\ (6.5)$ \\ \hline\hline
\end{tabular}
\end{table*}

%%% Systematics
Systematic uncertainties on $R$ associated with the fit
are determined by running ensembles of fits in which
input parameters are distributed according to our knowledge of the
underlying source, and include the PDF parametrization
(2\% to 12\%); the composition of combinatoric
backgrounds (2\% to 11\%); the mixture of $D^{**}$ states in
$B\to D^{**}\ellm\nulb$ decays ($0.4$\% to 6\%); the $B\to\Dstar\ellm\nulb$
form factors (0.1\% to 1.8\%); and the \piz efficiency, which affects the
$\Dstar\to D$ feed-down rate (0.4\% to 1.0\%). Uncertainties on
the \mmiss resolution for $B\to\Dstar\ellm\nulb$
events and on the $B\to D\ellm\nulb$ form factors each contribute less than 1\%.
The net systematic uncertainty on $R$ associated with fit yields is given by
$(\Delta R/R)_\mathrm{fit}$ for each channel in Table~\ref{tab:results}.
Uncertainties on $R$ propagated from the ratio of efficiencies for 
signal and normalization modes are typically small due to cancellations,
and include the limited statistics in the simulation ($1.1\%$ to $1.5\%$)
and systematic errors related to detector performance. The latter 
are determined by studying the efficiency
of track and neutral reconstruction and particle identification performance in
control samples in data and contribute less than $0.2\%$ each, except for
$\epm$ and $\mu^\pm$ identification, which contribute $0.5\%$ to $0.7\%$ each, and
are larger because the lepton momentum spectrum differs between
the signal and normalization processes. Finally, the uncertainty on 
$\cal B(\taum\to\ellm\nulb\nut)$~\cite{PDG} contributes $0.2\%$ to all modes.
The net systematic uncertainty on $R$ due to the efficiencies is given by
$(\Delta R/R)_{\varepsilon}$ in Table~\ref{tab:results}. 

These results are preliminary. We estimate that
uncertainties in $R$ due to modeling of bremsstrahlung radiation are
at or below the 1\% level, but they have not been explicitly included in
the results. While $B\to\ds\ellm\nulb$ decays
are modeled with HQET-based form factors~\cite{CLN}
including recent experimental measurements~\cite{FF},
we currently use ISGW2~\cite{ISGW2} to model signal decays.

We determine the statistical significance of the signals from $\sqrt{2\Delta(\ln\cal L)}$,
where $\Delta(\ln\cal L)$ is the change in log-likelihood between the nominal fit and
the no-signal hypothesis. The total significance is determined in a similar manner, by
modifying the likelihood function to take into account systematic uncertainties
from the fit. Table~\ref{tab:results} gives both significances for each channel.

%%% Conclusions
We have presented preliminary measurements of the decays
$B\to D\taum\nutb$ and $B\to\Dstar\taum\nutb$, relative to the
corresponding decays involving light leptons. We obtain
$R(B\to D\taum\nutb)=(40.7\pm 12.0\pm 4.9)\%$ and
$R(B\to\Dstar\taum\nutb)=(31.0\pm 5.7\pm 1.8)\%$,
where the first error is statistical and the second is systematic.
Normalizing to known branching fractions~\cite{PDG}, we obtain

\begin{eqnarray}
{\cal B}(B\to D\taum\nutb)&=&(0.90\pm 0.26\pm 0.11\pm 0.06)\% \nonumber\\
{\cal B}(B\to \Dstar\taum\nutb)&=&(1.81\pm 0.33\pm 0.11\pm 0.06)\%,\nonumber
\end{eqnarray}

\noindent where the third error is the uncertainty on the normalization
branching fraction, and where results are expressed for the \Bm lifetime.
The significances of the signals are $3.5\sigma$ and $6.2\sigma$, respectively.
The measurement of $B\to \Dstar\taum\nutb$ is consistent with a preliminary Belle measurement~\cite{Belle};
the measurement of $B\to D\taum\nutb$ is the first evidence for this mode.
These results are about $1\sigma$ higher than predictions based on the Standard Model, but, given the
size of the uncertainty, there is still room for a non-SM contribution. 

%%% Input the pubboard acknowledgements file
We are grateful for the excellent luminosity and machine conditions
provided by our \pep2\ colleagues, 
and for the substantial dedicated effort from
the computing organizations that support \babar.
The collaborating institutions wish to thank 
SLAC for its support and kind hospitality. 
This work is supported by
DOE
and NSF (USA),
NSERC (Canada),
CEA and
CNRS-IN2P3
(France),
BMBF and DFG
(Germany),
INFN (Italy),
FOM (The Netherlands),
NFR (Norway),
MIST (Russia),
MEC (Spain), and
STFC (United Kingdom). 
Individuals have received support from the
Marie Curie EIF (European Union) and
the A.~P.~Sloan Foundation.


\begin{thebibliography}{99}
\bibitem{Ell}Throughout this letter, we use the symbol $\ell$ to refer only to the light charged
  leptons $e$ and $\mu$. The symbol $D^{(*)}$ refers either to a $D$ or a $D^*$ meson.
\bibitem{ChargeConjugate} Throughout this letter, charge-conjugate modes are implied.
\bibitem{Isospin} This constraint follows from isospin symmetry in both the signal
and normalization modes, but is, in fact, slightly more general.
\bibitem{NormBF} We use~\cite{PDG} to normalize the four individual branching fractions. For
the \Bm--\Bzb-constrained measurement, we use our own averages of the values in~\cite{PDG}:
${\cal B}(\Bm\to\Dz\ellm\nulb)=(2.21\pm 0.15)\%$ and
${\cal B}(\Bm\to\Dstarz\ellm\nulb)=(5.85\pm 0.20)\%$.
\end{thebibliography}
\end{document}